%% file: main.tex
\documentclass[runningheads,a4paper]{llncs}

\input{macros}

\begin{document}
\title{Measuring Responsibility in \\ Multi-Agent Systems}

\author{Chunyan Mu \and Nir Oren}
\institute{Department of Computing Science\\ University of Aberdeen \\
\mailsa }

\maketitle

\begin{abstract}
We introduce a family of quantitative measures of responsibility in multi-agent planning, building upon the concepts of causal responsibility proposed by Parker et al.~\cite{ParkerGL23}. These concepts are formalised within a variant of probabilistic alternating-time temporal logic. Unlike existing approaches, our framework ascribes responsibility to agents for a given outcome by linking probabilities between behaviours and responsibility through three metrics, including an entropy-based measurement of responsibility. This latter measure is the first to capture the causal responsibility properties of outcomes over time, offering an asymptotic measurement that reflects the difficulty of achieving these outcomes. Our approach provides a fresh understanding of responsibility in multi-agent systems, illuminating both the qualitative and quantitative aspects of agents' roles in achieving or preventing outcomes.
\end{abstract}

\input{1-intro}
\input{2-model}

\input{3-cr-spec}

\input{4-cr-degree}
\input{5-concl}

\bibliographystyle{splncs03}
\bibliography{BIB-resp}


\input{6-appendix}

\end{document}

%% file: macros.tex
\usepackage{enumerate,amsmath,amssymb,graphicx}
\usepackage{url}
\usepackage{fancyvrb}
\usepackage{graphicx}
\usepackage{epsfig}
\usepackage{tikz,xspace,multicol,pgfplots}
\usepackage{soul}

\usepackage{enumitem}

\usepackage{algorithm}
\usepackage{algpseudocode}

\usepackage{comment}

\usetikzlibrary{positioning,shapes,arrows,automata}
\tikzset{ state/.style={draw,ellipse,initial text=} }

\tikzset{
  loop above right/.style={above right, out= 60, in= 30, loop},
  loop above left/.style ={above left,  out=150, in=120, loop},
  loop below right/.style={below right, out=330, in=300, loop},
  loop below left/.style ={below left,  out=240, in=210, loop}
}

\newcommand{\rpatl}	{$\gamma$ATL\,}

\newcommand{\ie}    	{i.e., }
\newcommand{\eg}    	{e.g., }

\newcommand{\true}   	{{\mathtt{true}}}

\newcommand{\prob}   	{{\sf {Prob}}} 
\newcommand{\last}   	{\mathit{last}}
\newcommand{\dist}   	{{\sf{Dist}}}

\def\A{\mathbf{A}}
\def\F{\Diamond} 
\def\G{\Box} 
\def\P{\mathbf{P}}
\def\M{\mathbf{M}}
\def\H{\mathbf{H}}

\def\U{\mathbf{U}}

\def\X{\bigcirc}

\renewcommand{\vec}[1]{\boldsymbol{\mathrm{#1}}}

\newcommand{\HIST}[2]{\mathsf{Hist}_{#2}(#1)}
\newcommand{\TRACE}[2]{\mathsf{Traces}_{#2}(#1)}
\newcommand{\PLAN}[3]{\mathsf{Plan}^{#3}_{#2}(#1)}

\newcommand{\ACT}{\mathsf{Act}}
\newcommand{\AG}{\mathsf{Ag}}
\newcommand{\ASP}{\mathsf{Ap}}

\newcommand{\CAR} {\mathsf{CAR}}
\newcommand{\CPR} {\mathsf{CPR}}
\newcommand{\CCR} {\mathsf{CCR}}

\newcommand{\coop} {\texttt{\scriptsize Coop}}
\newcommand{\defect} {\texttt{\scriptsize Defect}}

\newcommand{\cooperative} {\textit{\footnotesize co-op}}
\newcommand{\reward} {\textit{\footnotesize reward}}
\newcommand{\fine} {\textit{\footnotesize fine}}
\newcommand{\payoff} {\textit{\footnotesize payoff}}

\newcommand{\FS}{\rightarrow}
\newcommand{\CAL}[1]{\mathcal{#1}}

\newcommand{\DDD}{\CAL{D}}

\newcommand{\GGG}{\CAL{G}}
\newcommand{\HHH}{\CAL{H}}
\newcommand{\LLL}{\CAL{L}}

\newcommand{\CCC}{\mathbb{C}}
\newcommand{\PPP}{\mathbb{P}}

\newcommand{\TRANS}[1]{\stackrel{#1}{\longrightarrow}}
\newenvironment{GRAMMAR}{\[\begin{array}{lcl}}{\end{array}\]}
\newcommand{\VERTICAL}{\  \mid\hspace{-3.0pt}\mid \ }

\definecolor{blue}{rgb}{0,0,1}

\urldef{\mailsa}\path|chunyan.mu@abdn.ac.uk, nir.oren@abdn.ac.uk|

%% file: 1-intro.tex
\section{Introduction}
\label{sec:intro}

In Multi-Agent Systems (MASs), existing approaches to responsibility analysis typically provide a binary assessment as to whether an agent is responsible for a specific outcome. 
Such analysis is useful in ensuring that the MAS operate effectively, responsibly, and ethically across various domains. However, in many cases, responsibility is not binary. Instead, an agent can be partially responsible for an outcome, and as MASs grow in complexity, a more fine-grained (and quantitative) understanding of responsibility becomes increasingly important. In contrast to existing approaches, we introduce \emph{quantitative responsibility} analysis, seeking to provide a numerical measure signifying the extent to which an agent bears responsibility for some outcome.

While existing works have extensively explored aspects of responsibility in multi-agent systems \cite{ChocklerH04,BrahamVH12,AlechinaHL17,YazdanpanahDJAL19,ParkerGL23,BaierBK0P24}, our focus lies in the quantitative analysis of responsibility within the context of coalition plans. This framework allows for a more \textit{precise} understanding and anticipation of overall responsibility for outcomes with time. 
Specifically, in alignment with the notions proposed in \cite{ParkerGL23}, we focus on probabilistic and entropy-based measures of causal responsibility (CR) for certain outcomes and their associated concepts, namely Causal Active Responsibility (CAR), Causal Passive Responsibility (CPR), and Causal Contributive Responsibility (CCR). A CAR measure identifies the degree of responsibility the agent bears for bringing about some outcome, CPR measures the effort required by the agent to avoid an outcome, and CCR measures how much the agent contributes to an outcome across all agents in the system.

Outcomes are described in a variant of probabilistic Alternating-time Temporal Logic (ATL) \cite{AlurHK02,ChenL07} over finite traces, extended to include multiple CR operators. 
ATL is an expressive logic for reasoning about complex temporal MAS properties, and is highly modular. Taking advantage of this modularity, we extend the logic with new operators to formalise CR in a quantitative manner. This extension facilitates a unified framework capable of integrating quantitative measures into the logical structure, offering a comprehensive (quantitative) analysis of responsibility.

\paragraph{Related work and our contributions.}

There has been a significant  amount of research around the analysis of responsibility within MASs. Given the vast size of the literature, we focus only on those most closely related to this paper.

Alechina et al. \cite{AlechinaHL17} explored responsibility and blame in the context of team plan failures, focusing on determining causality and responsibility based on the definitions introduced by Halpern \cite{Halpern15}. \cite{YazdanpanahDJAL19,Baier0M21a,Baier0M21b,GladyshevADD23} explored the distribution of responsibility to individual agents and groups of agents within a formal verification framework, while \cite{BaierBK0P24} took a different angle, assigning a responsibility value to each state in a transition system that does not satisfy a given property.
Building on the former work, as well as on \cite{HalpernK18,FriedenbergH19}, Parker et al.~\cite{ParkerGL23} explored responsibility in multi-agent systems via definitions for active, passive and contributive responsibility, and through the use of structural equation modelling. Braham and van Hees \cite{BrahamVH12} studied responsibility attribution within a game-theoretic framework, emphasising causal responsibility's role in determining outcomes. Naumov and Tao \cite{NaumovT21} addressed responsibility in imperfect information strategic games, introducing blameworthiness and employing logical operators to  formalise these concepts. While  integrating agent strategies into their definitions of responsibility, less emphasis is placed on  different types of causal responsibility. 

In contrast to the work mentioned so far our approach formalises strategies, plans, and property specifications by extending probabilistic ATL. The extended logic, \rpatl, introduces a comprehensive quantitative dimension to responsibility analysis, particularly regarding the degrees of CR for outcomes with timing considerations. Specifically, our work advances responsibility analysis through three diverse metrics: behaviour counting, probability, and information entropy. Our entropy-based measure is the first to capture responsibility properties of outcomes over time, providing an asymptotic measurement that reflects the difficulty of achieving outcomes. We extend formal languages and provide systematic algorithms within the model checking framework. Thus, we  offer a refined and comprehensive analysis of responsibility in multi-agent planning scenarios beyond existing qualitative approaches.

%% file: 2-model.tex
\section{Model}
\label{sec:model}

This section presents a formal model to facilitate the structured analysis of responsibility in stochastic MASs.  

\subsection{The Stochastic Model of MASs}
The model of MASs is based on stochastic game structure with finite traces.
\begin{definition}
\label{def:model}
A \emph{multi-agent stochastic transition system} is a tuple $\GGG=(\AG, S, \ACT, \delta, 2^{\ASP}, L)$, where
\begin{itemize}

\item $\AG=\{1, \dots, n\}$ is a finite set of \emph{agents};

\item $S$ is a finite non-empty set of \emph{states};


\item $\ACT = \{a_1, a_2, \dots \}$ is a non-empty finite set of actions;

\item $\delta: S \times \ACT^{\AG} \FS \dist(S)$ is the \emph{probabilistic transition function} mapping from a state and joint actions to a distribution over new states;

\item $\ASP$ is a finite set of \emph{atomic propositions}; 

\item $L: S \FS 2^{\ASP}$ is the \emph{state labelling function} mapping each state to a set of atomic proposition drawn from $\ASP$.

\end{itemize}
\end{definition}
\noindent
We write $(s, \alpha) \to \mu$ for $s\in S$, $\alpha \in \ACT^{\AG}$, $\mu \in \delta(s, \alpha)$,
and write $s \TRANS{\alpha} s'$ whenever $(s,\alpha) \in \mu$ and $\mu(s')>0$.
\begin{definition}
\label{def:history}
A \emph{history} $\rho$ (\ie a finite path) is a non-empty finite sequence $s_0 \alpha_0 s_1 \alpha_1 \dots s_k$ of states and joint actions, 
where $\alpha_i \in \ACT^{\AG}$ is the $i^{th}$ joint action,
$s_{i+1} \in \delta(s_i,\alpha_i)$, meaning that the next state $s_{i+1}$ is drawn from the distribution defined by the transition function $\delta$ for state-action pair $(s_i,\alpha_i)$ for every $i$.
Let $\rho_s(i)$ denote the $i^{th}$ state of $\rho$, and $\rho_{\alpha}(i)$ denote the $i^{th}$ joint action of $\rho$, so for all $i$, we have $\rho_s(i)\TRANS{\rho_{\alpha}(i)} \rho_s(i+1)$.
Let $\HIST{s}{\GGG}$ denote the set of histories starting from state $s$, which we refer to as \emph{$\GGG$-histories}.
\end{definition}
Given $\GGG$ and $\rho$ starting from $s \in S$,
the \emph{cone} generated by $\rho$, written $\langle \rho \rangle$, is the set of complete paths ending in a terminal state, i.e.,
$\langle \rho \rangle = \{\rho' \in \HIST{s}{\GGG} \mid \rho \le \rho'\}$,
where the ordering notation $\le$ denotes the prefix relation. Given such a system and a state $s \in S$, 
we can then calculate the probability value, denoted by $\PPP_s(\rho)$, 
of any history $\rho$ starting at $s$ as follows:
\begin{itemize}
\item $\PPP_s(s) = 1$, and
\item $\PPP_s(\rho \TRANS{\alpha} s') = \PPP_s(\rho) \mu(s')$ for
$(\last(\rho), \alpha) \FS \mu$.
In other words, for the last joint action in $\rho$ from $\rho$'s final state $\last(\rho)$ which leads to $s'$. 
\end{itemize}
Let $\Omega_s \triangleq \HIST{s}{\GGG}$ be the sample space, 
and let $\GGG_s$ be the smallest $\sigma$-algebra induced by the cones 
generated by all the finite paths of $\GGG$. 
Then $\PPP$ induces a unique \emph{probabilistic measure} on $\GGG_s$ 
such that $\PPP_s(\langle \rho \rangle) = \PPP_s(\rho)$ 
for every history $\rho$ starting in $s$. 
\begin{definition}
\label{def:trace}
The \emph{trace} of a history $\rho = s_0, \alpha_0, s_1, \alpha_1 \dots s_k$ is the sequence of joint actions obtained by erasing the states, yielding action sequence $tr(\rho) =\alpha_0 \alpha_1 \dots \alpha_k$. 
We denote the set of traces for a system $\GGG$ starting from $s$ as $\TRACE{s}{\GGG}$.
\end{definition}
\begin{definition}
\label{def:strategy}
A \emph{strategy} of $i \in \AG$ is a map from the histories starting from the current state $s$
to a probability distribution over the agent's set of actions $\ACT$: 
$\sigma_i \triangleq \HIST{s}{\GGG} \FS \dist(\ACT)$, the set of all possible strategies for $i$ is denoted by $\Sigma_i$.
A \emph{joint strategy} or \emph{strategy profile} is a tuple of \emph{strategies} for a set of agents: $\vec{\sigma}=(\sigma_1, \sigma_2, \dots, \sigma_n)$, also represented as: $\HIST{s}{\GGG} \FS \dist(\ACT^{\AG})$ \ie assigning a distribution over the set of joint actions for each history $\rho$.
\end{definition}
\begin{definition}
\label{def:outcome}
A history generated by a strategy profile $\vec{\sigma}$ starting with the initial state $s_0$ is said to be consistent with the strategy profile, and is denoted by $\rho(s_0,\vec{\sigma})$, which can be expressed as: $\rho_s(0) = s_0$, $\rho_s(k+1) \in \delta(\rho_s(k), \alpha_k)$.The set of consistent  histories generated by $\vec{\sigma}$ starting from $s_0$ is denoted by $\HIST{s_0, \vec{\sigma}}{\GGG}$, and the associated traces are denoted by $\TRACE{s_0, \vec{\sigma}}{\GGG}$.
\end{definition}
\begin{definition}
\label{def:plan}
Given a non-empty coalition of agents $J \subseteq \AG$ and starting state $s \in S$, a \emph{joint plan} is a function $\pi: J \FS \TRACE{s, \vec{\sigma}_{J}}{\GGG}$, where $\vec{\sigma}_J$ denotes the strategy profile of the coalition $J$. $\HIST{s}{\pi}$ is the history consistent with $\pi$ from $s$. 
\end{definition}
\begin{definition}
\label{def:j-compatible}
Two joint plans $\pi_1$ and $\pi_2$ are  $\langle J \rangle$-compatible if the actions taken by coalition $J \subseteq \AG$ along the histories consistent with $\pi_1$ and $\pi_2$ are equivalent, denoted as $\pi_1 \sim_{\langle J \rangle} \pi_2$. 
$\PLAN{s}{\pi}{\langle J \rangle}$ is the equivalence class of plans which are $\langle J \rangle$-compatible with $\pi$ starting from state $s$.
\end{definition}
\begin{example}
\label{eg:model}
The Continuous Prisoners' Dilemma (CPD) scenario \cite{WahlN99} allows agents to repeatedly interact by undertaking `cooperate' or `defect' actions, and is illustrated in  Figure \ref{fig:eg}. If both agents cooperate,  both receive a reward; if one cooperates, the cooperator receives a (negative) payoff; if both defect, they both receive a fine.
\input{fig-cpd-model}
Each agent can act by either performing a `\coop' or `\defect' action (cooperating or defecting  respectively), possible joint actions are: 
$\alpha_0=\defect_1 \defect_2$,
$\alpha_1=\coop_1 \coop_2$,
$\alpha_2=\coop_1 \defect_2$,
$\alpha_3=\defect_1 \coop_2$. 
If agent $i$ executes a $\coop$ action, then proposition $\cooperative_i$ is true in the resultant state. Possible states are thus:
\begin{eqnarray*}
s_0 &=&  \fine \triangleq \neg \cooperative_1 \land \neg \cooperative_2 \\
s_1 &=&  \reward \triangleq \cooperative_1 \land \cooperative_2 \\
s_2 &=& \payoff_1 \triangleq \cooperative_1 \land \neg \cooperative_2 \\
s_3 &=& \payoff_2 \triangleq \neg \cooperative_1 \land \cooperative_2
\end{eqnarray*}
Consider two joint plans $\pi_1,\pi_2$ when the game is played twice, and which result in the following  traces:
$\pi_1  = \alpha_2, \alpha_3 $, $\pi_2  = \alpha_1, \alpha_0 $.
Clearly, we have that $\pi_1 \sim_{\{A_1\}} \pi_2$ since the actions taken by $A_1$ along the two plans are the same.
\end{example}

\subsection{Probabilistic ATL}
ATL \cite{AlurHK02} extends traditional temporal logics, allowing for the expression of properties related to concurrent decision-making and strategic interactions among agents. 
Probabilistic ATL (pATL, \eg \cite{ChenL07}) builds upon ATL by introducing a probabilistic dimension, enabling the modelling of uncertainty and randomness in multi-agent systems. We adapt probabilistic ATL to express properties over finite paths, aligning it with our objective of (quantitative) responsibility analysis.
%
\begin{definition}[Syntax]
\label{def:syntax}                                                         
Let $\GGG = (\AG, S, \delta, \ASP, L)$.
The \emph{syntax} of pATL is made up of
\emph{state formulae} and
\emph{history formulae}
represented by $\phi$ and $\psi$ respectively.
\begin{GRAMMAR}
 \phi
     &::=&
  a 
     \VERTICAL
  \neg \phi
     \VERTICAL
  \phi \land \phi
     \VERTICAL
  \langle A \rangle \lbrack \psi \rbrack
     \VERTICAL
  \P_{\bowtie p} \, \langle A \rangle \lbrack {\psi} \rbrack
     \\
  \psi, \varphi
     &::=&
  \X \phi 
     \VERTICAL
  \phi  \U_{\le k} \phi 
    \VERTICAL
  \neg \psi
     \VERTICAL
  \psi \land \varphi     
\end{GRAMMAR}
\noindent Here $a \in \ASP$ is an \emph{atomic proposition},
$A \subseteq \AG$ is a set of agents,
$\langle A \rangle \lbrack \psi \rbrack$ expresses the property that coalition $A$ has a joint strategy to enforce $\psi$,
${\bowtie} \in \{\le, <, \ge, >\}$, 
$k \in \{0,1,2,...\}$ is a time bound,
and $p \in \lbrack 0,1 \rbrack$ is a probability bound.  
\end{definition}
\noindent
The formula $\P_{\bowtie p} \ \langle A \rangle \lbrack \psi \rbrack$ expresses that coalition $A$ has a strategy such that the probability of satisfying history formula $\psi$ is $\bowtie p$, when the strategy is followed.
%
\begin{definition}[semantics] 
\label{def:semantics}
For a state $s \in S$ of $\GGG$, the \emph{satisfaction relation}
$s \models_{\GGG} \phi$ for state formulae denotes ``$s$ satisfies $\phi$'':

\begin{itemize}

  \item $s \models_{\GGG} a$ iff $a \in L(s)$.

  \item $s \models_{\GGG} \neg\phi$ iff $s \not \models_{\GGG} \phi$.

  \item $s \models_{\GGG} \phi \land \phi'$ iff 
  	$s \models_{\GGG} \phi$ and  $s \models_{\GGG}  \phi'$.

 \item $s \models_{\GGG} \langle A \rangle \lbrack \psi \rbrack$ iff \ 
	$\exists \vec{\sigma}_{A}$ s.t. 
	$\forall \rho \in \HIST{s, \vec{\sigma}_A}{\GGG}. \rho \models_{\GGG} \psi$.
	
	
 \item $s \models_{\GGG} \P_{\bowtie p} \langle A \rangle \lbrack \psi \rbrack$ iff \ 
	$\exists \vec{\sigma}_{A}$, s.t. 
	$\prob(s, [\![\langle A \rangle \lbrack \psi \rbrack ]\!]) \bowtie p$,
	where: 
	$$\prob(s, [\![ \langle A \rangle \lbrack \psi \rbrack ]\!])  
	= \PPP_s\{\rho \in \HIST{s,\vec{\sigma}_{A}}{\GGG} \mid 
	\rho \models_{\GGG} \psi \}.$$

\end{itemize}

\noindent 
For a history $\rho \in \HIST{s_0}{\GGG}$, the \emph{satisfaction relation}
$h \models_{\GGG} \varphi$ for history formulae denotes ``$h$ satisfies $\varphi$'':

\begin{itemize} 

\item $\rho \models_{\GGG} \X \phi$ iff $\rho_s(1) \models \phi$.
  
\item $\rho \models_{\GGG} \phi \U_{\le k} \phi'$ iff $\exists i \le k.\rho_s(i) \models_{\GGG} \phi'$ and $\forall j <i. \rho_s(j) \models_{\GGG} \phi$.

\item $\rho \models_{\GGG} \neg\psi$ iff $\rho \not \models_{\GGG} \psi$.

\item $\rho \models_{\GGG} \psi \land \varphi$ iff $\rho \models_{\GGG} \psi$ and  $\rho \models_{\GGG}  \varphi$.
  
\end{itemize}

\end{definition}
The until operator $\U$ allows one to derive the temporal modalities $\F$ (``eventually'') and $\G$ (``always''): $\F_{\le k} \psi  \triangleq  \true~ \U_{\le k}~ \psi$ and $\G_{\le k} \psi  \triangleq  \neg \F_{\le k} (\neg\psi)$.

%% file: fig-cpd-model.tex
\begin{figure} 
\begin{minipage}{0.58\linewidth}
\begin{eqnarray*}
\AG &=& \{A_1, A_2\} \\
\ACT &=& \{\coop, \defect\} \\
\ASP &=& \{\cooperative_1, \cooperative_2\} \\
S &=& \{s_0, s_1, s_2, s_3\} \\
\end{eqnarray*}
\end{minipage}\hfill
\begin{minipage}{0.4\linewidth}
  \begin{tikzpicture}[->,auto,node distance=1.6cm, thick,main node/.style={circle,draw,font=\sffamily\bfseries}]
      \node[main node] (1) {$s_0$};
      \node[main node]  [right of=1](2) {$s_1$};
      \node[main node] [below of=1](3) {$s_2$};
      \node[main node] [below of=2] (4) {$s_3$};
      \path (1) edge[<->] node[pos=0.2]{\scriptsize$\alpha_0$} node[below,pos=0.8]{\scriptsize$\alpha_1$}  (2);
      \path (1) edge[<->] node[left,pos=0.3]{\scriptsize$\alpha_0$} node[pos=0.8]{\scriptsize$\alpha_2$} (3);
      \path (1) edge[<->] node[left,pos=0.3]{\scriptsize$\alpha_0$} node[pos=0.8]{\scriptsize$\alpha_3$} (4);
      \path (2) edge[<->]   (3);
      \path (2) edge[<->] node[left,pos=0.3]{\scriptsize$\alpha_1$} node[pos=0.8]{\scriptsize$\alpha_3$} (4);
      \path (3) edge[<->] node[below,pos=0.3]{\scriptsize$\alpha_2$} node[pos=0.8]{\scriptsize$\alpha_3$} (4);
      \path(1) edge[loop above left] node {\scriptsize $\alpha_0$} ();
      \path(2) edge[loop above right] node {\scriptsize $\alpha_1$} () ;
      \path(3) edge[loop below left] node {\scriptsize $\alpha_2$} () ;
      \path(4) edge[loop below right] node {\scriptsize $\alpha_3$} () ; 
    \end{tikzpicture}
\end{minipage}
\caption{The transition system representing CPD.} 
\label{fig:eg}
\end{figure}
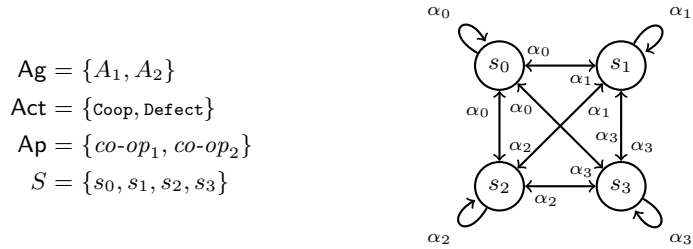   

%% file: 3-cr-spec.tex
\section{Formalising Responsibility}
\label{sec:cr-spec}

We incorporate the causal responsibility definitions introduced in~\cite{ParkerGL23} into our model and extend the logic of pATL into a formulation we refer to as \rpatl by adding the following elements to state formulae. 
{\small \begin{GRAMMAR}
 \phi
     &::=&
  \dots
     \VERTICAL
  \CAR_{\GGG}(i, \pi, \varphi)
     \VERTICAL
  \CPR_{\GGG}(i, \pi, \varphi)
     \VERTICAL
  \CCR_{\GGG}(i, \pi, \varphi)    
\end{GRAMMAR}}
%
\begin{definition}[Causal Active Responsibility (CAR)]
\label{def:car}
Given $\GGG$, $i \in \AG$, a joint plan $\pi$, and an outcome $\varphi$,
we say $i$ bears CAR  for $\varphi$ in $\pi$ at $s$, 
denoted via the operator $\CAR_{\GGG}(i, \pi, \varphi)$ in \rpatl, 
if $\varphi$ holds for all possible histories consistent with $\PLAN{s}{\pi}{\langle \{i\} \rangle}$ while $\varphi$ does not hold in some histories consistent with $\PLAN{s}{\GGG}{\langle \AG \rangle}$. 
We define the semantics for the operator $\CAR_{\GGG}(i, \pi, \varphi)$ in \rpatl as follows:
\begin{eqnarray*}
s &\models_{\GGG}& \CAR_{\GGG}(i, \pi, \varphi) \text{ iff } \\
&& \forall \pi' \in \PLAN{s}{\pi}{\langle \{i\} \rangle} . \forall \rho' \in \HIST{s}{\pi'}. \rho' \models_{\GGG} \varphi \ \land \\
&& \exists \pi'' \in \PLAN{s}{\GGG}{\langle \AG \rangle} . \forall \rho'' \in \HIST{s}{\pi''}. \rho'' \not\models_{\GGG} \varphi
\end{eqnarray*}
\end{definition}
Intuitively, agent $i$ takes active responsibility for the occurrence of the outcome expressed by $\varphi$ in $\pi$ starting at $s_0$ if keeping  $i$'s actions fixed the other agents could not avoid the outcome by choosing different actions. 
A brief algorithm for checking the CAR operator is presented in Algorithm \ref{alg:car}\footnote{Algorithms to instantiate our framework are not central to understanding our approach, and therefore are provided in the supplementary material.}. 
\begin{example}
\label{eg:car}
Consider the outcome expressed by $$\varphi = \langle A_1, A_2 \rangle \X (\fine \lor \payoff_2)$$ in Example~\ref{eg:model}, and  a joint plan:
$\pi = (\defect_1 \defect_2)$. 
Note that $\forall \pi' \in \PLAN{s_0}{\pi}{\langle \{A_1\} \rangle}$, $\HIST{s_0}{\pi'} \models_{\GGG} \varphi$, (\ie for $\pi' = (\defect_1 \coop_2)$ or $\pi'=(\defect_1 \defect_2)$). Therefore, while keeping the initial state and actions of $A_1$ fixed, the other agents ($A_2$ in this example) could not have acted differently to avoid the occurrence of $\varphi$. Now consider $\pi''=(\coop_1 \defect_2)$, and note that $\HIST{s}{\pi''} \not\models_{\GGG} \varphi$, \ie there exists a joint plan $\pi''\in \PLAN{s_0}{\GGG}{\langle \{A_1,A_2\} \rangle}$ which avoids the occurrence of $\varphi$. Thus, $s_0 \models_{\GGG} \CAR_{\GGG}(A_1,\pi,\varphi)$ and $A_1$ bears CAR  for $\varphi$ in $\pi$ at $s_0$.
\end{example}
\begin{definition}[Causal Passive Responsibility (CPR)]
\label{def:cpr}
Given $\GGG$, $i \in \AG$, a joint plan $\pi$, and an outcome $\varphi$, 
we say $i$ bears CPR  for $\varphi$ in $\pi$ at $s$, denoted by the operator $\CPR_{\GGG}(i, \pi, \varphi)$ in \rpatl, 
if $\varphi$ holds for all possible histories consistent with $\pi$, 
while violated in some histories consistent with $\PLAN{s}{\pi}{\langle \AG \setminus \{i\}\rangle}$.
We define the semantics for the operator $\CPR_{\GGG}(i, \pi, \varphi)$ in \rpatl as follows:
\begin{eqnarray*}
s &\models_{\GGG}& \CPR_{\GGG}(i, \pi, \varphi) \text{ iff } \
 \forall \rho \in \HIST{s}{\pi}. \rho \models_{\GGG} \varphi \ \land \\
&& \exists \pi' \in \PLAN{s}{\pi}{\langle \AG \setminus \{i\} \rangle} \text{ s.t.\ }  
\forall \rho' \in \HIST{s}{\pi'}. \rho' \not\models_{\GGG} \varphi
\end{eqnarray*}
\end{definition}
Intuitively, an agent $i$ takes passive responsibility for the occurrence of outcome expressed in $\varphi$ in $\pi$ starting at $s$ if keeping all other agents' actions fixed agent $i$ could avoid the outcome by choosing different actions. The algorithm for checking the CPR operator is outlined in Algorithm ~\ref{alg:cpr}.
\begin{example}
\label{eg:cpr}
Consider the outcome expressed by $$\varphi =\langle A_1, A_2 \rangle \X \reward$$ in Example~\ref{eg:model}, 
given a joint plan $\pi = (\coop_1 \coop_2)$. 
Here, $\HIST{s_0}{\pi} \models \varphi$.
Consider $\pi' = (\defect_1 \coop_2)$, and 
note that $\pi' \in \PLAN{s_0}{\pi}{\langle \{A_2\} \rangle}$ and $\HIST{s_0}{\pi'} \not\models \varphi$,
\ie by keeping the initial state and the actions of all other agents fixed, $A_1$ can instead execute action $\coop_1$ to  avoid the realisation of $\varphi$. Thus, $A_1$ bears CPR  for $\varphi$ in $\pi$ at $s_0$.
\end{example}
\begin{definition}[Causal Contributive Responsibility (CCR)]
\label{def:ccr}
Given $\GGG$, $i \in \AG$, a joint plan $\pi$, and an outcome $\varphi$,
we say $i$ bears CCR  for $\varphi$ in $\pi$ at $s$, denoted by the operator $\CCR_{\GGG}(i, \pi, \varphi)$ in \rpatl, 
if $\varphi$ holds for all possible histories consistent with $\pi$, 
and there exists some coalition of agents $J \subseteq \AG$, and agent $i \in J$ such that 
all histories consistent with all compatible plans in $\PLAN{s}{\GGG}{\langle J \rangle}$ satisfy $\varphi$, while there exists compatible plans in $\PLAN{s}{\pi}{\langle J \setminus \{i\} \rangle}$ for which all possible consistent histories violate $\varphi$.
We define the semantics for the operator $\CCR_{\GGG}(i, \pi, \varphi)$ in \rpatl as follows:
{\small \begin{eqnarray*}
s &\models_{\GGG}& \CCR_{\GGG}(i, \pi, \varphi)  
\text{ iff } \ (\forall \rho \in \HIST{s}{\pi}, \rho \models_{\GGG} \varphi) \land \\
&& \exists J \subseteq \AG \mbox{ s.t. } ((i \in J) \land \\
&&  (\forall \pi' \in \PLAN{s}{\pi}{\langle J \rangle} .  
\forall \rho' \in \HIST{s}{\pi'}. \rho' \models_{\GGG} \varphi)   \land  \\
&&  (\exists \pi'' \in \PLAN{s}{\pi}{\langle J \setminus \{i\} \rangle} . 
\forall \rho'' \in \HIST{s}{\pi''}. \rho'' \not\models_{\GGG} \varphi))
\end{eqnarray*}}
\end{definition}
Intuitively, an agent $i$ takes contributive responsibility for the occurrence of an outcome in a joint plan starting at $s$ if $i$ is a part of a coalition $J$ such that the actions of $J$ are sufficient to lead to the outcome while the actions of $J$ without $i$ are insufficient to lead to the outcome. The algorithm for checking the CCR operator is presented in Algorithm~\ref{alg:ccr}.
\begin{example}
\label{eg:ccr}
Continue Example~\ref{eg:model}, consider the outcome expressed by $\varphi = \langle A_1, A_2 \rangle \F_{\le 2} \ \fine$,  
given a joint plan: $\pi {=} (\coop_1 \defect_2), (\defect_1 \defect_2)$. 
Here, $\HIST{s_0}{\pi} \models \varphi$.
Now consider $J=\{A_1,A_2\}$, 
and notice that $\PLAN{s}{\pi}{\langle {A_1,A_2} \rangle} = \{\pi\}$, indicating that all actions of $J$ are needed to ensure the occurrence of $\varphi$. If we consider $\pi'' = (\coop_1 \defect_2), (\coop_1 \defect_2)$, we see
that $\pi'' \in \PLAN{s}{\pi}{\langle \{A_2\} \rangle}$ and $\HIST{s}{\pi''}\not\models \varphi$. Thus,  the actions of $J \setminus {A_1}$ are inadequate to ensure the occurrence of $\varphi$, and
so $A_1$ bears (CCR) for $\varphi$ in $\pi$ at $s_0$.
\end{example}
%
%
%
Turning to the complexity of checking for CAR, CPR and CCR, recall that the complexity of model checking ATL and pATL is in $P^{\text{NP} \cap \text{coNP}}$ \cite{ChatterjeeAH06,ChenL07}. 
This leads to the following theorem which follows directly from our algorithms.
\begin{theorem}
\label{theo:complex1}
The complexity of checking CAR and CPR is $M \cdot P^{NP \cap co-NP}$, and $M \cdot 2^{|Ag|} \cdot P^{NP \cap co-NP}$ for CCR. Here, $M$ is the number of histories and $|Ag|$ is the number of agents in the system.


\begin{proof}
In the worst-case scenarios, the algorithms for checking CAR, CPR, and CCR exhaustively consider all possible plans and consistent histories satisfying the formula $\varphi$. 

For CAR and CPR, the algorithms involve nested loops that iterate over all possible compatible plans under different conditions. For each plan, the algorithms further iterate over all consistent histories, performing model checking for the satisfaction of $\varphi$ in each iteration. For CCR, there is an additional loop over all subsets of $\AG$ outside the nested loops.

Given model checking ATL and pATL formulae is $P^{NP \cap co-NP}$ \cite{ChatterjeeAH06,ChenL07}, the overall time complexity of the responsibility checking algorithms is bounded by $M \cdot P^{NP \cap co-NP}$ for CAR and CPR, and $M \cdot 2^{|Ag|} \cdot P^{NP \cap co-NP}$ for CCR. 
\hfill $\Box$
 \end{proof}
\end{theorem}
This result indicates that checking CCR is more computationally intensive than checking for CAR and CPR. Intuitively this is due to the need to evaluate responsibility across all possible coalitions, resulting in another exponential term relative to the number of agents.

%% file: 4-cr-degree.tex
\section{Measuring Causal Responsibility}
\label{sec:cr-degrees}
We seek to identify the \emph{level of responsibility an agent bears for a given outcome in a joint plan}. We quantify the degree of responsibility based on three distinct metrics: proportion, probability, and information entropy of relevant behaviours. 

\subsection{Metrics} \label{sec:metrics}
We introduce our concept for measuring behaviours within the context of a general language.
Given a finite system $\GGG$ over alphabet $\Sigma$ (\ie the joint actions $\ACT^{\AG}$), we denote its language by $\LLL(\GGG)$.
For a language $\LLL\subseteq \Sigma^*$, $\LLL^n=\LLL \cap \Sigma^n $ denotes the set of length $n$ of words in $\LLL$.

\paragraph{Proportional measure.}

Perhaps the simplest approach to measuring behaviours involves counting the cardinality of elements which satisfy the behaviour. We denote by $\CCC(\LLL_{\varphi}) = |\LLL_{\varphi}|$ the number of elements (\ie histories) in our language which satisfy $\varphi$. We refer to this measure as the \emph{proportional measure}. Such a measure is useful in scenarios where one wants to assess the impact of a specific strategy on the outcome. By counting possible behaviours regarding what might have happened if a different strategy had been taken, proportional measures provide insights into the causal relationships between strategies and outcomes. 

\paragraph{Probabilistic measure.}
An alternative approach to quantification involves computing the probabilities associated with behaviours leading to the realisation of the outcome. Given $\GGG$ and $s_0 \in S$, the probability value, denoted as $\PPP_s(\rho)$, for any history $\rho \in \HIST{s}{\pi}$ generated by $\pi$ starting at $s_0$ can be calculated as described earlier.
Probabilistic measures  are particularly useful in scenarios where outcomes are influenced by probabilistic factors. This approach allows for a more complex and precise assessment of the likelihood of different behaviours. 
While probabilistic measures are often relevant, they have a significant limitation: such measures can be too precise. For many interesting properties the probability (on very long or infinite behaviours) is either 0 or 1, making quantitative analysis infeasible. 
\begin{example}
\label{eg:entropy-1}
Consider the scenario presented in Example~\ref{eg:model}, 
and the properties: $\psi = \Box_{\le t}(\fine \lor \reward)$.
and $\psi' = \Box_{\le t} \Diamond_{<100} (\payoff_1 \lor \payoff_2)$.
Intuitively, when $t$ is large, an agent must undertake a certain level of effort to ensure that $\psi$ is satisfied (as it has to be maintained across all $t$ units of time), while less effort is required to satisfy $\psi'$ (as the condition must hold at least once every 100 time units).
However, a probability analysis for both properties over very long runs tends towards 0, and therefore disagrees with this intuition: 
{\small \begin{eqnarray*}
\P(\langle A_1, A_2 \rangle \psi) &=& \lim_{t \to \infty} (\frac{1}{2})^{t} \to 0 \\ 
\P(\langle A_1, A_2 \rangle \psi') &=& \lim_{t \to \infty} (1-\frac{1}{2^{100}})^{t} \to 0 
\end{eqnarray*}}
\end{example}
\noindent
Nevertheless, it seems counter-intuitive to say that satisfying $\psi'$ is as challenging as satisfying $\psi$.
%
\paragraph{Entropy measure.}
Given this weakness of probabilistic measures, we turn our attention to measurement based on the notion of \emph{entropy} as an alternative. This approach offers a measure of the information content in the system. In situations where traditional probability measures might lack granularity, entropy provides a more refined view, capturing the complexity and uncertainty in the system.
The \emph{entropy} of a language (of finite words)  $\LLL \subseteq\Sigma^*$ (\cite{Chomsky58}) is defined as:
$$
\HHH(\LLL) = \limsup_{n \to \infty} \frac{\log_2 (1+|\LLL^n|)}{n}
$$
Intuitively, the entropy of a language represents the quantity of information (measured in bits per symbol) contained in ``typical'' words of that language.
For a regular language $\LLL$ over the alphabet $\Sigma$ accepted by a finite automaton, the computation of its entropy is computationally achievable. Specifically, given a deterministic finite automaton (DFA) $\A$, we define $\A$ as \emph{trimmed} if all its states are reachable from the initial state and co-reachable to a final state. Let $\M(\A)$ represent its \emph{extended} adjacency matrix, where $\M(\A)_{ij}$ counts the number of symbols $a \in \Sigma$ such that the DFA permits the transition.
The computation of entropy is obtained thanks to the following theorem.
\begin{theorem} [\cite{Chomsky58}]\label{thm:rho}
For any finite deterministic trimmed automaton $\A$,
$$
\HHH (\LLL(\A)) = \log \varrho(\M(\A)),
$$
where $\varrho(\M)$ denotes the spectral radius of matrix $\M$ (\ie the maximum modulus of its eigenvalues).
\end{theorem}
\noindent
In general, 
if $\LLL = \HIST{s}{\varphi}$, where $\HIST{s}{\varphi}$ are the histories of $\GGG$ satisfying formula $\varphi$, and $|\Sigma|=k$ then $\HHH(\LLL) = \log k$. 
%
%
\begin{example}
\label{eg:entropy-2}
Consider again the situation discussed in Example \ref{eg:entropy-1}. 
We can calculate the entropy of the system: $\HHH(\GGG)= \log 4 = 2$.
The entropy for paths satisfying $\psi$ and $\psi'$ is approximately 
$\HHH(\GGG \cap \psi) \approx 1$ and
$\HHH(\GGG \cap \psi') \approx 1.99$, respectively.
This aligns with our intuition, suggesting that, on average, almost half of the system behaviours involve visiting states $s_0$ and $s_1$ in each transition step. This reflects the effort needed to guide the system to satisfy $\psi$ by eliminating all  behaviours leading to $s_2$ and $s_3$. Almost all system behaviors visit the $s_2$ and $s_3$ within every 100 steps, indicating little effort is required to guide the system to satisfy $\psi'$.
\end{example}
In MAS planning scenarios requiring a comprehensive assessment of responsibility, one can leverage a combination of these measures to different ends: proportional measures provide insights into overall behaviour ratios and causal impact; probability measures effectively handle uncertainty, offering precise evaluations; and entropy-based measures deal with information content over extended periods of time. This approach facilitates a comprehensive analysis, taking into account causality, uncertainty considerations, and information dynamics among agents.

\subsection{Measuring causal responsibilities}
We can now measure the degree of responsibility ascribed to an agent using the described metrics.
\begin{definition}[CAR degrees]
Given $\GGG$, $i \in \AG$, a joint plan $\pi$, and an outcome $\varphi$. 
Let:
{\small \begin{eqnarray*}
\LLL^{+}_{\CAR(i,\pi,\varphi)} = \{\rho \in \HIST{s_0}{\pi'} \mid \rho \models \varphi \ \text{and} \ \pi' \in \PLAN{s_0}{\pi}{\langle \{i\} \rangle}\} &&\\
\LLL^{-}_{\CAR(i,\pi,\varphi)}  = \{\rho \in \HIST{s_0}{\pi''}  \mid \rho \not\models \varphi \ \text{and} \ \pi'' \in \PLAN{s_0}{\GGG}{\langle \AG \rangle}\} &&
\end{eqnarray*}} 
Then: $\DDD^{X}_{\CAR}(i, \pi, \varphi)= Y(\LLL^{+}_{\CAR(i,\pi,\varphi)})/Y(\LLL_{\varphi})\cdot \kappa^-$. If $X=\#$ and $Y=\CCC$ then the degree of CAR is computed via a  proportional measure; if $X=\P$ and $Y=\PPP$ then the degree of CAR is computed via a probabilistic measure; and if $X=\H$ and $Y=\HHH$ then it is calculated through an entropy measure. Here $\kappa^- = 1$ if $|\LLL^{-}_{\CAR(i,\pi,\varphi)} )|>0$ (i.e., is avoidable) and $\kappa^-=0$ otherwise.
\end{definition}
In essence, the degree of CAR for agent $i$ concerning  outcome $\varphi$ under plan $\pi$ is characterised by the extent to which  $i$ can influence the system to fulfil the specified property. This is quantified by the ratio of behaviours that adhere to $\varphi$ when following the strategies devised by $i$ within $\pi$, relative to all possible behaviours that lead to $\varphi$, when the occurrence of $\varphi$ is avoidable (\ie when $\kappa^- > 0$). 
The algorithm to compute the degrees of CAR underpinned by the metrics of Section \ref{sec:metrics} is sketched in Algorithm \ref{alg:dcar}.
\begin{example}
\label{eg:dcar}
Consider Example \ref{eg:car} with $i=A_1$, and assume each agent executes action $ \defect $ \ with probability $\frac{1}{4}$, and $\coop$ \ with probability $\frac{3}{4}$. 

(a) Consider outcomes and plan: 
$$\varphi = \langle A_1, A_2 \rangle \X (\fine \lor \payoff_2), \pi=(\defect_1,\coop_2)$$
we have:
{\small 
\begin{eqnarray*}
&& \LLL^{+}_{\CAR(A_1,\pi,\varphi)} = \{(\defect_1 \coop_2), (\defect_1 \defect_2) \}   \\
&& \LLL^{-}_{\CAR(A_1,\pi,\varphi)}  = \{(\coop_1 \defect_2), (\coop_1 \coop_2)\}  \\
&& \DDD^{\#}_{\CAR}(A_1, \pi, \varphi) = \DDD^{\P}_{\CAR}(A_1, \pi, \varphi) = \DDD^{\H}_{\CAR}(A_1, \pi, \varphi) = 1 
\end{eqnarray*}
}
\noindent Since $A_1$'s  $\defect_1$ in $\pi$ facilitates outcome $\varphi$, $A_1$ has full CAR for this outcome, which also meets our intuition. 

(b) Now consider outcome and plan:
$$\varphi= \langle A_1, A_2 \rangle \F_{\le t} (\reward \lor \payoff_2), \pi= \dots (\defect_1,\coop_2)$$
we have:
\noindent {\small 
\begin{eqnarray*}
&& \LLL^{+}_{\CAR(A_1,\pi,\varphi)} = \{\dots (\defect_1 \coop_2) \}  \\ 
&& \LLL^{-}_{\CAR(A_1,\pi,\varphi)}  = \{\dots (\coop_1 \defect_2), \dots (\defect_1 \defect_2)\} \\%
&& \DDD^{\#}_{\CAR}(A_1, \pi, \varphi) = \frac{3^{t-1}}{2*3^{t-1}} \cdot 1 = \tfrac{1}{2},  \\ 
&& \DDD^{\P}_{\CAR}(A_1, \pi, \varphi) = \frac{3/16*(1/4)^{t-1}}{(3/16+9/16)*(1/4)^{t-1}} \cdot 1 = \tfrac{1}{4}, \\
&& \DDD^{\H}_{\CAR}(A_1, \pi, \varphi) = \frac{\log (1+3^{t-1})}{\log (1+2*3^{t-1})} \cdot 1. 
\end{eqnarray*}}
\input{fig-dcar-eg}

\noindent Here, both the proportional and probability measures indicate that $A_1$ maintains a fixed partial degree of CAR, with the probability measure providing a more precise assessment due to the non-uniform distribution of actions. The entropy measure yields a more asymptotic degree measurement over time.
When $t$ is small, indicating a relatively harder-to-reach outcome, $A_1$'s active responsibility is relatively small. As $t$ increases, making the outcome progressively more accessible, $A_1$'s active responsibility increases. This aligns with intuition as larger values of $t$ correspond to easier attainment of the outcome and consequently greater active responsibility for $A_1$. 
Figure \ref{fig:car-eg-plot} plots CAR for the 3 measures.
\end{example}
\begin{definition}[CPR degrees]
Given $\GGG$, $i \in \AG$, a joint plan $\pi$, and an outcome  $\varphi$.
Let:
{\small $$\LLL^{+}_{\CPR(i,\pi,\varphi)} = \{\rho \in \HIST{s_0}{\pi'} \mid \rho \models \varphi \ \text{and} \ \pi' \in \PLAN{s_0}{\pi}{\langle \AG \rangle}\}$$
$$\LLL^{-}_{\CPR(i,\pi,\varphi)} = \{\rho \in \HIST{s_0}{\pi''}  \mid \rho \not\models \varphi \ \text{and} \ \pi'' \in \PLAN{s_0}{\pi}{\langle \AG \setminus \{i\} \rangle}\}$$}
Then:
$\DDD^{X}_{\CPR}(i, \pi, \varphi)=Y(\LLL^{-}_{\CPR(i,\pi,\varphi)})/Y(\LLL_{\neg\varphi})\cdot \kappa^+$.
If $X=\#$ and $Y=\CCC$ then the degree of CPR is obtained via a proportional measure; if $X=\P$ and $Y=\PPP$ then the degree of CPR  is computed via a probabilistic measure; and if $X=\H$ and $Y=\HHH$ then it is calculated via an entropy measure. Here $\kappa^+=1$ if $|\LLL^{+}_{\CPR(i,\pi,\varphi)} )|>0$ (i.e., is avoidable) and $\kappa^+=0$  otherwise.

\end{definition}
\noindent
Intuitively, the CPR degree of agent $i$ for the occurrence of $\varphi$ under the plan $\pi$ measures the level of difficulty for  $i$ to \emph{avoid} the outcome by selecting different actions while keeping all other agents' actions fixed. This is quantified by the ratio of behaviours in which $i$ varies its actions while other agents maintain compatible actions in $\pi$, relative to all possible behaviours violating $\varphi$, when the occurrence of $\varphi$ under the plan $\pi$ is achievable (\ie when $\kappa^+ > 0$). The algorithm to compute the degrees of CPR is provided in Algorithm \ref{alg:dcpr}.
\begin{example}
\label{eg:dcpr}
Continue Example \ref{eg:model}, consider $i=A_1$ and 
$\varphi = \langle A_1, A_2 \rangle \F_{\le t} \ \reward$
and assume each agent executes $\defect$ \ and $\coop$ \ with probability $1/2$ and $1/2$ (uniform) respectively. Consider $\pi=\dots (\coop_1,\coop_2)$, we have:
{\small \begin{eqnarray*}
\LLL^{+}_{\CPR(A_1,\pi,\varphi)} &=& \{\dots (\coop_1 \coop_2) \} \\
\LLL^{-}_{\CPR(A_1,\pi,\varphi')}  &=& \{\dots (\defect_1 \coop_2)\} 
\end{eqnarray*}}
{\small
\begin{align*}
\DDD^{\#}_{\CPR}(A_1, \pi, \varphi) = \DDD^{\P}_{\CPR}(A_1, \pi, \varphi)= \tfrac{1}{3} \\
\DDD^{\H}_{\CPR}(A_1, \pi, \varphi) = \frac{\log (1+3^{t-1})}{\log (1+3^t)}  
\end{align*}
}
\noindent 
Both proportional and probability measures indicate that $A_1$ bears a third of CPR. The entropy-based measure changes over time: as $t$ increases, indicating a smaller difficulty in achieving the outcome and a concomitant greater challenge to avoid the outcome, $A_1$'s passive responsibility along the given plan increases, aligning with intuition.
Figure \ref{fig:cpr-eg-plot} illustrates this example.
\input{fig-dcpr-eg}
On the other hand, assume we consider the joint plan $\pi=(\defect_1,\coop_2)^t$ instead, it can be observed that $i$ does not bear CPR, and all degrees of CPR are zero, as the outcome is unattainable following $\pi'$ (as $\kappa^+=0$).
\end{example}
\begin{definition}[CCR degrees]
Given $\GGG$, $i \in \AG$, a joint plan $\pi$, and an outcome $\varphi$ .
For all $J \subseteq \AG$, let:
{\small \begin{eqnarray*}
\LLL^{J,+}_{\CCR(i,\pi,\varphi)} = \{\rho \in \HIST{s_0}{\pi'} &\mid&  \rho \models_{\GGG} \varphi \ \text{and} \ \\
&& \pi' \in \PLAN{s_0}{\pi}{\langle J \rangle} \land i \in J\} \\
\LLL^{J,-}_{\CCR(i,\pi,\varphi)} = \{\rho \in \HIST{s_0}{\pi''} &\mid& \rho \not\models_{\GGG} \varphi \ \text{and} \ \\
&&  \pi'' \in \PLAN{s_0}{\pi}{\langle J \setminus \{i\} \rangle}\} 
\end{eqnarray*}} 
Then

$$\DDD^{X}_{\CCR}(i, \pi, \varphi)= \frac{\sum_{\forall J \subseteq \AG} Y(\LLL^{J,+}_{\CCR(i,\pi,\varphi)}) \cdot \kappa^{J,-} / Y(\LLL_{\varphi}) }{\#J.(|\LLL^{J,+}_{\CCR(i,\pi,\varphi)} )|>0)} $$

If $X=\#$ and $Y=\CCC$ then the degree of CPR is obtained via a proportional measure; if $X=\P$ and $Y=\PPP$ then the degree of CPR  is computed via a probabilistic measure; and if $X=\H$ and $Y=\HHH$ then it is calculated via an entropy measure. Here $\#J.\varphi$  represents the count of subsets $J$ for which $\varphi$ holds, and  $\kappa^{J,-} = 1$ if $|\LLL^{J,-}_{\CCR(i,\pi,\varphi)} )|>0$ (i.e., is avoidable); $\kappa^{J,-}=0$  otherwise. 

\end{definition}
\noindent
Intuitively, the CCR degree for agent $i$ for the occurrence of $\varphi$ under plan $\pi$, defines how much $i$ could \emph{contribute} to the occurrence of the outcome on average considering all possible $i$-included coalitions' joint plans that could lead to the outcome when it is unavoidable ($\kappa^{J,-}>0$). 
The correction factor $\kappa^-$ distinguishes between avoidable and unavoidable outcomes, adjusting the degree accordingly. The algorithm to compute the degrees of CCR is provided in Algorithm \ref{alg:dcpr}.

\begin{example}
\label{eg:dccr}
Returning to Example \ref{eg:ccr},
we assume each agent takes action $\defect$ and $\coop$ with uniform likelihood (\ie with probability $1/2$ for each action). 
Consider $$\varphi = \langle A_1, A_2 \rangle \G_{\leq t} \ (\fine \lor \reward)$$ 
where the joint plan $\pi$ is given by $(\text{catch}_1 , \text{catch}_2), (\text{skip}_1 , \text{skip}_2) \dots $. Thus, $A_1$ and $A_2$ consistently choose the same action at each step, alternating between $\text{catch}$ and $\text{skip}$, repeating this for $t$ steps. Here, $J$ can be either $\{A_1\}$ or $\{A_1,A_2\}$, each case requires that in each step of the corresponding joint plan $\pi''$ forming $\LLL^{-}_{\CCR(i,\pi,\varphi)}$, $A_1$ must consistently opt for a different action to $A_2$ to satisfy the conditions (violating $\varphi$) imposed by CCR. 
Note that in this scenario,
$$\LLL_{\varphi} = \{(\defect_1 \defect_2)^{k_1} (\coop_1 \coop_2)^{k_2} \mid k_1 + k_2 = t\}$$ 
so $\CCC (\LLL_{\varphi}) = 2^t,\ \PPP(\LLL_{\varphi})=1/2^t,\ \HHH(\LLL_{\varphi}) = \log 2$. Thus:

{\small \begin{align*}
\DDD^{\#}_{\CCR}(A_1, \pi, \varphi) & =\DDD^{\P}_{\CCR}(A_1, \pi, \varphi)=\tfrac{1}{2^t}\\
\DDD^{\H}_{\CCR}(A_1, \pi, \varphi) & = \tfrac{1}{2t}
\end{align*}}
\noindent
In scenarios with a small value of $t$, the realisation of $\varphi$ is relatively easy, and the degree of CCR that $A_1$ bears should be relatively significant. All measures indicate a positive degree of CCR for $A_1$, aligning with intuition. As $t$ increases, achieving $\varphi$ becomes more challenging, and the CCR measure should gradually decrease. All measures show a decrease, but the probability measure converges faster, while the entropy measure converges more slowly.
Intuitively, as $t$ grows, $A_1$ should partially bear CCR for the outcome, and this contribution should gradually diminish as $t \to \infty$. The entropy-based measure decays more slowly, and thus agrees better with this intuition. Figure \ref{fig:ccr-eg-plot} illustrates this visually.
\end{example}

\input{fig-dccr-eg}

\paragraph{Complexity.}
In quantitative cases, the complexity is slightly higher than in qualitative cases due to additional factors such as computing the cardinality or probability of each history, which is \emph{linear} with respect to the length of the histories. The complexity of computing the degree of CAR, CPR and CAR build directly on the logical properties of the operators, and are thus equivalent to the operator complexity as described in Theorem \ref{theo:complex1}.


\begin{proposition} The following simple properties of CR degrees can be readily obtained by their definitions.
\begin{itemize}
\item If $\LLL^-_{\CAR(i,\pi,\varphi)}=\{\}$  (c.f. $\LLL^+_{\CPR(i,\pi,\varphi)}$ and $\LLL^{J,-}_{\CCR(i,\pi,\varphi)}$ for all $J \subseteq \AG$), then all CAR (c.f. CPR and CCR) degree measures equal $0$.

\item If $|\LLL^+_{\CAR(i,\pi,\varphi)}| = |\LLL_{\varphi}|$ (c.f. $|\LLL^-_{\CPR(i,\pi,\varphi)}| = |\LLL_{\neg\varphi}|$ and $|\LLL^+_{\CAR(i,\pi,\varphi)}| = |\LLL_{\varphi}|$), then all CAR (c.f. CPR and CCR) degree measures equal $1$.
\end{itemize}
\end{proposition}


\begin{remark}[Proportional and probability measure]
Under a uniform distribution, the probabilistic measures of CAR, CPR, and CCR are equivalent to their respective proportional measures. This alignment simplifies the assessment of responsibility degrees, providing a more direct relationship between the probabilistic and proportional perspectives when considering uniform strategy distributions.
\end{remark}

\begin{remark}[Probability and entropy measure]
The probability-based measure evaluates an agent's responsibility by considering the likelihood of paths where the agent actively, passively, or contributively influences an outcome. It computes the ratio of the probability of paths where the agent is responsible to the total probability of paths leading to or avoiding the outcome. The entropy-based measure assesses responsibility by capturing the variability and growth rate of behaviours leading to or avoiding an outcome. It calculates the entropy of the set of paths where the agent is responsible, providing insights into the diversity of the agent's contributions over time.
\end{remark}

\begin{remark}[CCR and Shapely values \cite{Shapley52}] There are some conceptual similarities between the degree of CCR and the Shapely value. The latter assigns a value to players based on their marginal contributions to all possible coalitions, while the former assesses the contribution of an agent's responsibility towards an outcome by quantifying how much they contribute to it as part of a joint plan. These similarities suggest that further investigations as to the links between the two would be fruitful.
\end{remark}

In summary, our causal responsibility measures consider the \textit{nature} of the outcome (achievable/unachievable, avoidable/unavoidable) and incorporates it into the calculation, offering a \textit{context-aware} evaluation of an agent's active,  passive, and contributive responsibility. The measures are sensitive to variations in agent behaviour and demonstrate distinctions in responsibility degrees based on different joint plans and outcomes over \textit{time}. We believe the measures align with intuitive expectations of responsibility, providing reasonable and interpretable results in diverse multi-agent scenarios.

%% file: fig-dcar-eg.tex
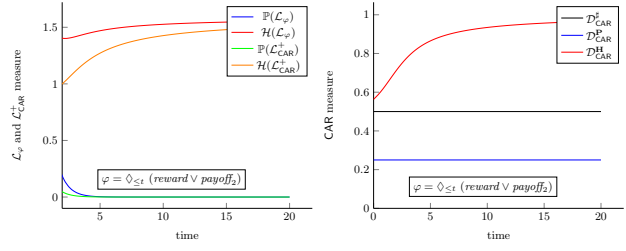
\begin{figure}
\centering

\scalebox{0.48}{\begin{tikzpicture}
\node[draw,align=right] at (3,0.8) {$\varphi=\F_{\le t} \ (\reward \lor \payoff_2)$};
\begin{axis}[
    ymin=-0.1,
    xmin=2,
    scaled ticks=false,
    tick label style={/pgf/number format/fixed},
    every axis plot post/.append style=
        {mark=none,domain=1:20,samples=50,thick,smooth},
    axis x line*=bottom,
    axis y line*=left,
    enlargelimits=upper,
    xlabel={time}, 
    ylabel={$\LLL_{\varphi}$ and $\LLL^+_{\CAR}$ measure},
    ]

    \addplot[blue] {12/16*(1/4)^(x-1)};
    \addplot[red] {(log2(1+2*3^(x-1))/x};
    \addplot[green] {3/16*(1/4)^(x-1)};
    \addplot[orange] {(log2(1+3^(x-1)))/x};
    \legend{$\PPP(\LLL_{\varphi})$, $\HHH(\LLL_{\varphi})$, $\PPP(\LLL^+_{\CAR})$, $\HHH(\LLL^+_{\CAR})$};
\end{axis}
\end{tikzpicture}}
\scalebox{0.48}{\begin{tikzpicture}
\node[draw,align=right] at (3,0.6) {$\varphi=\F_{\le t} \ (\reward \lor \payoff_2)$};
\begin{axis}[
    ymin=-0,
    scaled ticks=false,
    tick label style={/pgf/number format/fixed},
    every axis plot post/.append style=
        {mark=none,domain=0:20,samples=50,thick,smooth},
    axis x line*=bottom,
    axis y line*=left,
    enlargelimits=upper,
    xlabel={time}, 
    ylabel={$\CAR$ measure},
    ]

    \addplot[black] {0.5};
    \addplot[blue] {0.25};
    \addplot[red] {(log2(1+3^(x-1)))/(log2(1+2*3^(x-1))};
    \legend{$\DDD^{\sharp}_{\CAR}$, $\DDD^{\P}_{\CAR}$,  $\DDD^{\H}_{\CAR}$};
\end{axis}
\end{tikzpicture}}
\caption{$\CAR$ for the measures over time for  Example \ref{eg:dcar} (b)}
\label{fig:car-eg-plot}
\end{figure}

%% file: fig-dcpr-eg.tex
\begin{figure}
\centering
\scalebox{0.48}{\begin{tikzpicture}
\node[draw,align=left] at (3,1.2) {$\varphi=\F_{\le t} \ \reward$ \\ $\neg\varphi=\G_{\le y} \ \neg \reward$};
\begin{axis}[
    ymin=0,
    scaled ticks=false,
    tick label style={/pgf/number format/fixed},
    every axis plot post/.append style=
        {mark=none,domain=1:20,samples=50,thick,smooth},
    axis x line*=bottom,
    axis y line*=left,
    enlargelimits=upper,
    xlabel={time}, 
    ylabel={$\LLL_{\neg\varphi}$ and $\LLL^-_{\CPR}$ measure},
    ]

    \addplot[blue] {3/4};
    \addplot[red] {(log2(3^(x)+1)/(x)};
    \addplot[green] {1/4};
    \addplot[orange] {(log2(3^(x-1)+1)/(x))};
    \legend{$\PPP(\LLL_{\neg\varphi})$, $\HHH(\LLL_{\neg\varphi})$, $\PPP(\LLL^-_{\CPR})$, $\HHH(\LLL^-_{\CPR})$};
\end{axis}
\end{tikzpicture}}
\scalebox{0.48}{\begin{tikzpicture}
\node[draw,align=left] at (3,1) {$\varphi=\F_{\le t} \ \reward$ \\ $\neg\varphi=\G_{\le y} \ \neg \reward$};
\begin{axis}[
    ymin=0,
    scaled ticks=false,
    tick label style={/pgf/number format/fixed},
    every axis plot post/.append style=
        {mark=none,domain=1:20,samples=50,thick,smooth},
    axis x line*=bottom,
    axis y line*=left,
    enlargelimits=upper,
    xlabel={time}, 
    ylabel={$\CAR$ measure},
    ]

    \addplot[blue] {1/3};
    \addplot[red] {(log2(3^(x-1)+1))/log2(3^(x)+1)};
    \legend{$\DDD^{\P}_{\CPR}$, $\DDD^{\H}_{\CPR}$};
\end{axis}
\end{tikzpicture}}
\caption{$\CPR$ over time for Example \ref{eg:dcpr}}
\label{fig:cpr-eg-plot}
\end{figure}
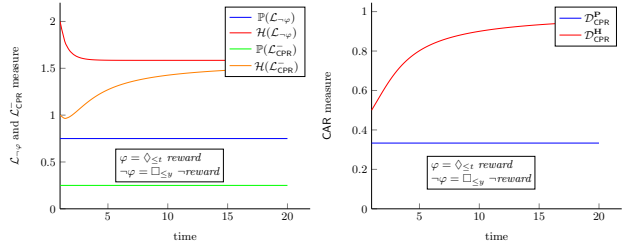

%% file: fig-dccr-eg.tex
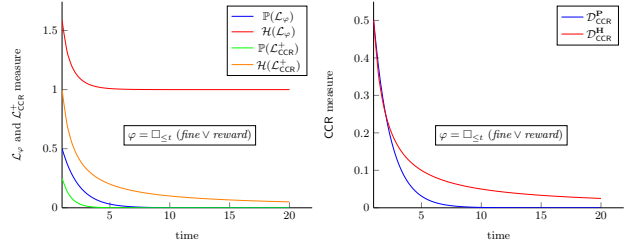
\begin{figure}
\centering
\scalebox{0.48}{\begin{tikzpicture}
\node[draw,align=left] at (3.5,2) {$\varphi=\G_{\le t} \ (\fine \lor \reward)$};
\begin{axis}[
    scaled ticks=false,
    tick label style={/pgf/number format/fixed},
    every axis plot post/.append style=
        {mark=none,domain=1:20,samples=50,thick,smooth},
    axis x line*=bottom,
    axis y line*=left,
    enlargelimits=upper,
    xlabel={time}, 
    ylabel={$\LLL_{\varphi}$ and $\LLL^+_{\CCR}$ measure},
    ]

    \addplot[blue] {(0.5)^x};
    \addplot[red] {log2(2^(x)+1)/x};
    \addplot[green] {(0.25)^x};
    \addplot[orange] {1/x};
    \legend{$\PPP(\LLL_{\varphi})$, $\HHH(\LLL_{\varphi})$, $\PPP(\LLL^+_{\CCR})$, $\HHH(\LLL^+_{\CCR})$};
\end{axis}
\end{tikzpicture}}
\scalebox{0.48}{\begin{tikzpicture}
\node[draw,align=left] at (3.5,2) {$\varphi=\G_{\le t} \ (\fine \lor \reward)$};
\begin{axis}[
    scaled ticks=false,
    tick label style={/pgf/number format/fixed},
    every axis plot post/.append style=
        {mark=none,domain=1:20,samples=50,thick,smooth},
    axis x line*=bottom,
    axis y line*=left,
    enlargelimits=upper,
    xlabel={time}, 
    ylabel={$\CCR$ measure},
    ]

    \addplot[blue] {0.5^x};
    \addplot[red] {1/(2*x)};
    \legend{$\DDD^{\P}_{\CCR}$, $\DDD^{\H}_{\CCR}$};
\end{axis}
\end{tikzpicture}}
\caption{$\CCR$ over time for Example \ref{eg:dccr}}
\label{fig:ccr-eg-plot}
\end{figure}

%% file: 5-concl.tex
\section{Conclusions and Future Work}
\label{sec:conc}

This paper explored the notion of quantitative causal responsibility within a MAS setting. To do this, we extended ATL with new operators aimed at formalising causal responsibility concepts and demonstrated that different metrics for such responsibility may be useful in different contexts. 

We have identified several avenues for future work including reasoning about tradeoffs between responsibility and coalition performance.
We also plan to incorporate strategic logic \cite{ChatterjeeHP10,AminofKMMR19} into our system to allow for the integration of Shapely values as discussed earlier. 
Finally, we will investigate the expressive power of \rpatl. Given that our newly introduced operators quantify over plans, histories and agents, we believe that \rpatl is more expressive than pATL, but will seek to determine how it relates to other logics such as PSL.


%% file: 6-appendix.tex
\section*{Appendix}

We have provided algorithms to instantiate our framework. These are not central to understanding our approach, and therefore appear in the appendix.

\input{alg-car}

\input{alg-cpr}

\input{alg-ccr}

\input{alg-dcar}

\input{alg-dcpr}

\input{alg-others}

\input{alg-dccr}

%
%
%

%% file: alg-car.tex
\begin{algorithm}
\caption{Checking $s \models_{\GGG} \CAR_{\GGG}(i, \pi, \varphi)$}
\begin{algorithmic}[1]
\State \textbf{Input:} $\GGG, s, i, \pi, \varphi$
    \For{each $\pi'$ in compatible plans $\PLAN{s}{\pi}{\langle\{i\}\rangle}$}
        \For{each consistent history $\rho \in \HIST{s}{\pi'}$}
            \If{$\rho \not\models_{\GGG} \psi$}
                \State \textbf{return} False
            \EndIf
        \EndFor
    \EndFor

    \For{each $\pi''$ in all possible plans $\PLAN{s}{\GGG}{\AG}$}
        \State eFlag $\gets$ True 
        \For{each consistent history $\rho \in \HIST{s}{\pi''}$}
            \If{$\rho \models_{\GGG} \psi$}
                \State eFlag $\gets$ False
                \State \textbf{break}
            \EndIf
        \EndFor
        \If{eFlag = True}
            \State \textbf{return} True
        \EndIf
    \EndFor
    
    \State \textbf{return} False
\end{algorithmic}
\label{alg:car}
\end{algorithm}

%% file: alg-cpr.tex
\begin{algorithm}
\caption{Checking $s \models_{\GGG} \CPR_{\GGG}(i, \pi, \varphi)$}
\begin{algorithmic}[1]
\State \textbf{Input:} $\GGG, s, i, \pi, \varphi$
    \For{each $\rho$ in $\HIST{s}{\pi}$}
        \If{$\rho \not\models_{\GGG} \varphi$}
            \State \textbf{return} False
        \EndIf
    \EndFor
    
    \For{each $\pi'$ in $\PLAN{s}{\pi}{\langle \AG \setminus \{i\} \rangle}$}
        \State eFlag $\gets$ True 
        \For{each $\rho'$ in $\HIST{s}{\pi'}$}
            \If{$\rho \models_{\GGG} \varphi$}
                \State eFlag $\gets$ False
                \State \textbf{break}
            \EndIf
        \EndFor
        \If{eFlag = True}
            \State \textbf{return} True
        \EndIf
    \EndFor
    
    \State \textbf{return} False
\end{algorithmic}
\label{alg:cpr}
\end{algorithm}

%% file: alg-ccr.tex
\begin{algorithm}
\caption{Checking $s \models_{\GGG} \CCR_{\GGG}(i, \pi, \varphi)$}
\begin{algorithmic}[1]
\State \textbf{Input:} $\GGG, s, i, \pi, \varphi$
\For{each $\rho$ in $\HIST{s}{\pi}$}
    \If{$\neg(\rho \models_{\GGG} \varphi)$}
        \State \textbf{return} False
    \EndIf
\EndFor

\For{each $J \subseteq \AG$ such that $i \in J$}
    \State $\text{allFlag} \gets \text{True}$ 
    \For{each $\pi' \in \PLAN{s}{\pi}{\langle J \rangle}$}
        \For{each $\rho \in \HIST{s}{\pi'}$}
            \If{$\rho \not\models_{\GGG} \varphi$}
                \State $\text{allFlag} \gets \text{False}$
                \State \textbf{break}
            \EndIf
        \EndFor
    \EndFor
    
    \If{allFlag=True}
        \For{each $\pi'' \in \PLAN{s}{\pi}{\langle J \setminus \{i\} \rangle}$}
            \State $\text{existFlag} \gets \text{True}$ 
            \For{each $\rho \in \HIST{s}{\pi''}$}
                \If{$\rho \models_{\GGG} \varphi$}
                    \State $\text{existFlag} \gets \text{False}$
                    \State \textbf{break}
                \EndIf
            \EndFor
            \If{existFlag=True}
                \State \textbf{return} True
            \EndIf
        \EndFor
    \EndIf
\EndFor

\State \textbf{return} False
\end{algorithmic}
\label{alg:ccr}
\end{algorithm}

%% file: alg-dcar.tex
\begin{algorithm}
\caption{Calculate $\DDD^X_{\CAR}(i, \pi, \varphi)$ for $X=\#, \P, \H$}
\begin{algorithmic}[1]
\State \textbf{Input:} $\GGG, s, i, \pi, \varphi$
\State \textbf{Output:} $\DDD^{\sharp}_{\CAR}(i, \pi, \varphi), \DDD^P_{\CAR}(i, \pi, \varphi), \DDD^H_{\CAR}(i, \pi, \varphi)$
    \State $c,p,h \gets 0,0,0$
    \State $\kappa \gets 0$
    \State $\LLL_{\varphi}, \LLL^+, \LLL^- \gets \{\}, \{\}, \{\}$
    
    \For{each history $\rho \in \HIST{s}{\GGG}$}
        \If{$\rho \models_{\GGG} \varphi$}
             \State $\LLL_{\varphi} \gets \LLL_{\varphi} \cup \{\rho\}$
        \EndIf
    \EndFor
    
    \For{each $\pi'$ in compatible plans $\PLAN{s}{\pi}{\langle\{i\}\rangle}$}
        \For{each consistent history $\rho \in \HIST{s}{\pi'}$}
            \If{$\rho \models_{\GGG} \varphi$}
                \State $\LLL^+ \gets \LLL^+ \cup \{\rho\}$
            \EndIf
        \EndFor
    \EndFor
    
    \For{each $\pi''$ in all possible plans $\PLAN{s}{\GGG}{\AG}$}
        \For{each consistent history $\rho \in \HIST{s}{\pi''}$}
            \If{$\rho \not\models_{\GGG} \varphi$}                
                \State $\LLL^- \gets \LLL^- \cup \{\rho\}$
            \EndIf
        \EndFor
    \EndFor

    \If{$|\LLL^-| > 0$} \State $\kappa \gets 1$ \EndIf

    \State $c \gets \CCC(\LLL^+)/\CCC(\LLL_{\varphi}) * \kappa$ \Comment{Alg. \ref{alg:card}}
    \State $p \gets \PPP(\LLL^+)/\PPP(\LLL_{\varphi}) * \kappa$ \Comment{Alg. \ref{alg:card}}
    \State $h \gets \log_2 c / \textsc{MaxLen}(\LLL_{\varphi}) * \kappa$ \Comment{Alg. \ref{alg:maxlen}}
    
    \State \textbf{Return} $(c,p,h)$
\end{algorithmic}
\label{alg:dcar}
\end{algorithm}

%% file: alg-dcpr.tex
\begin{algorithm}
\caption{Calculate $\DDD^X_{\CPR}(i, \pi, \varphi)$ for $X=\#, \P, \H$}
\begin{algorithmic}[1]
\State \textbf{Input:} $\GGG, s, i, \pi, \varphi$
\State \textbf{Output:} $\DDD^{\sharp}_{\CPR}(i, \pi, \varphi), \DDD^P_{\CPR}(i, \pi, \varphi), \DDD^H_{\CPR}(i, \pi, \varphi)$
    \State $c,p,h \gets 0,0,0$, $\kappa \gets 0$
    \State $\LLL_{\neg\varphi}, \LLL^+, \LLL^- \gets \{\}, \{\}, \{\}$
    
    \For{each history $\rho \in \HIST{s}{\GGG}$}
        \If{$\rho \not \models_{\GGG} \varphi$}
            \State $\LLL_{\neg\varphi} \gets \LLL_{\neg\varphi} \cup \{\rho\}$
        \EndIf
    \EndFor
    
    \For{each $\pi'$ in compatible plans $\PLAN{s}{\pi}{\langle\{\AG\}\rangle}$}
        \For{each consistent history $\rho \in \HIST{s}{\pi'}$}
            \If{$\rho \models_{\GGG} \varphi$}
                \State $\LLL^+ \gets \LLL^+ \cup \{\rho\}$
            \EndIf
        \EndFor
    \EndFor
    
    \For{each $\pi''$ in all possible plans $\PLAN{s}{\pi}{\AG \setminus \{i\}}$}
        \For{each consistent history $\rho \in \HIST{s}{\pi''}$}
            \If{$\rho \not\models_{\GGG} \varphi$}
                \State $\LLL^- \gets \LLL^- \cup \{\rho\}$
            \EndIf
        \EndFor
    \EndFor

    \If{$|\LLL^+| > 0$} \State $\kappa \gets 1$ \EndIf

    \State $c \gets \CCC(\LLL^-)/\CCC(\LLL_{\neg\varphi}) * \kappa$ \Comment{Alg. \ref{alg:card}}
    \State $p \gets \PPP(\LLL^-)/\PPP(\LLL_{\neg\varphi}) * \kappa$ \Comment{Alg. \ref{alg:card}}
    \State $h \gets \log_2 c / \textsc{MaxLen}(\LLL_{\neg\varphi}) * \kappa$ \Comment{Alg. \ref{alg:maxlen}}
    
    \State \textbf{Return} $(c,p,h)$
\end{algorithmic}
\label{alg:dcpr}
\end{algorithm}

%% file: alg-others.tex
\begin{algorithm}
\caption{Compute $\CCC(L)$ and $\PPP(L)$ for given language $L$}
\begin{algorithmic}[1]
\State \textbf{Input:} $\GGG, s_0, L$
\State \textbf{Output:} $\text{cardinality}(L), \text{probability}(L)$
    \State Initialise $\text{card} \leftarrow 0$
    \State Initialise $\text{prob} \leftarrow 0$

    \For{each word $w$ in $L$}
      \State $\text{card} \leftarrow \text{card} + 1$ 
      \State $\text{prob} \leftarrow \text{prob} + \textsc{Prob}(\GGG,s_0,w)$ \Comment{Alg. \ref{alg:prob}}
    \EndFor

    \State \textbf{return} $\text{card}$, $\text{prob}$
  \end{algorithmic}
\label{alg:card}
\end{algorithm}

\begin{algorithm}
  \caption{Compute probability of a word}
  \begin{algorithmic}[1]
    \Function{Prob}{$\GGG, s_0, \rho$}
      \State $s \gets s_0$ \Comment{Initialise the current state}
      \State $\rho \gets \text{empty path}$ \Comment{Initialise the path}
      \State $p \gets 1$ \Comment{Initialise probability}
      
      \ForAll{$(\alpha, s') \in \rho$} \Comment{Iterate over actions}
        \State $\rho \gets \rho \TRANS{\alpha} s'$ \Comment{Extend the path}
        \State $p \gets p \cdot \mu(s, \alpha, s')$ \Comment{Update probability}
        \State $s \gets s'$ \Comment{Update the current state}
      \EndFor
      
      \State \Return $p$
    \EndFunction
  \end{algorithmic}
\label{alg:prob}
\end{algorithm}

\begin{algorithm}[h!]
  \caption{Compute the maximum length of words in a language $L$}
  \begin{algorithmic}[1]
    \Function{MaxLen}{$L$}
        \State Initialise $\text{max\_length} \leftarrow 0$

        \For{each word $w$ in $L$}
        \State $\text{length} \leftarrow |w|$
        \If{$\text{length} > \text{max\_length}$}
            \State $\text{max\_length} \leftarrow \text{length}$
        \EndIf
        \EndFor

        \State \textbf{return} $\text{max\_length}$
    \EndFunction
  \end{algorithmic}
\label{alg:maxlen}
\end{algorithm}

%% file: alg-dccr.tex
\begin{algorithm}
\caption{Calculate $\DDD^X_{\CCR}(i, \pi, \varphi)$ for $X=\#, \P, \H$}
\begin{algorithmic}[1]
\State \textbf{Input:} $\GGG, s, i, \pi, \varphi$
\State \textbf{Output:} $\DDD^{\sharp}_{\CCR}(i, \pi, \varphi), \DDD^P_{\CCR}(i, \pi, \varphi), \DDD^H_{\CCR}(i, \pi, \varphi)$

\State $c,p,h \gets 0,0,0$, $\text{numJ} \gets 0$, $\kappa \gets 0$
\State $\LLL_{\varphi}, \LLL^+, \LLL^- \gets \{\}, \{\}, \{\}$
    
\For{each history $\rho \in \HIST{s}{\GGG}$}
    \If{$\rho \models_{\GGG} \varphi$}
        \State $\LLL_{\varphi} \gets \LLL_{\varphi} \cup \{\rho\}$
    \EndIf
\EndFor

\State $xJ \gets \{\}$
\For{each $J \subseteq \AG$ such that $i \in J$}
    \For{each $\pi' \in \PLAN{s}{\pi}{\langle J \rangle}$}
        \For{each $\rho \in \HIST{s}{\pi'}$}
            \If{$\rho \models_{\GGG} \varphi$}
                \State $\LLL^+ \gets \LLL^+ \cup \{\rho\}$
            \EndIf
        \EndFor
    \EndFor

    \For{each $\pi'' \in \PLAN{s}{\pi}{\langle J \setminus \{i\} \rangle}$}
        \For{each $\rho \in \HIST{s}{\pi''}$}
            \If{$\rho \not\models_{\GGG} \varphi$}
                \State $\LLL^- \gets \LLL^- \cup \{\rho\}$
            \EndIf
        \EndFor
    \EndFor

    \If{$|\LLL^-| > 0$} 
        \State $xJ \gets xJ \cup \{J\}$
        \State $\text{numJ} \gets \text{numJ}+1$ 
        \State $c \gets c+\CCC(\LLL^+)$ \Comment{Alg. \ref{alg:card}}
        \State $p \gets p+\PPP(\LLL^+)$ \Comment{Alg. \ref{alg:card}}
        \State $h \gets h+\log_2 c $ \Comment{Alg. \ref{alg:maxlen}}
    \EndIf
\EndFor

\If{numJ $> 0$}
    \State $c \gets c /\CCC(\LLL_{\varphi}) /\text{numJ} $ \Comment{Alg. \ref{alg:card}}
    \State $p \gets p /\PPP(\LLL_{\varphi}) /\text{numJ} $ \Comment{Alg. \ref{alg:card}}
    \State $h \gets h / \textsc{MaxLen}(\LLL_{\varphi}) / \text{numJ} $ \Comment{Alg. \ref{alg:maxlen}}
\EndIf

\State \textbf{Return} $(c,p,h)$
\end{algorithmic}
\label{alg:dccr}
\end{algorithm}